\documentstyle[12pt]{article}
\def\be{\begin{equation}}
\def\ee{\end{equation}}
\def\bea{\begin{eqnarray}}
\def\eea{\end{eqnarray}}

\begin{document}

\begin{center}
{\Large{\bf Investigation of Zero-Modes for a Dynamical
D$p$-Brane}}

\vskip .5cm {\large Farzin Safarzadeh-Maleki } \vskip .1cm
{\sl f.safarzadeh@aut.ac.ir}\\
\end{center}

\begin{abstract}

In this article, we investigate zero-modes for a dynamical
(rotating-moving) D$p$-brane, coupled to the electromagnetic and
tachyonic background fields. This work is done by the boundary
state methods, in three cases of bosonic and fermionic boundary
states and superstring partition function. By analyzing the
obtained zero-modes in either of the cases, interesting results
will be obtained. Our findings demonstrate the importance of the
zero-mode and its effects on the background fields and the
defined internal properties of the described system.

\end{abstract}


 \vskip .5cm

\section{Introduction and Conclusion}

So far many significant aspects of D-branes \cite {1,2}, as an
essential objects of string/superstring theory, have been
discovered. D-branes, interpreted as the classical solutions of
the low energy string effective action, could be defined in terms
of closed strings. Meanwhile, the boundary state method \cite
{3,4}, as a powerful technique, can be considered to show the
couplings of all closed string states to D-branes. Boundary state
method is a beneficial approach in many complicated situations,
even when a clear space-time is not accessible. By applying the
boundary state method for describing D-branes, different
properties and many configurations of these objects have been
studied \cite {5}-\cite {7}. Another description of D-branes, as
tachyonic solitons, follows from the boundary string field
theory, in which one can codify the information by using the disc
partition function of the open string sigma model. In this
context, for the case of superstring theory, an effective
space-time action can be obtained by the corresponding
world-sheet partition function \cite {8}-\cite {10}.

In most articles in which have been studied the issue of the
boundary state, the zero-modes have been neglected for
simplifying the calculation and hence, the zero-modes effects
have been omitted from the system. In this paper, we concentrate
specifically on the zero-modes and their effects on the
properties of the system under study, such as its dynamics and
background fields. This work is done by studying the zero-modes
boundary states, for both bosonic and fermionic theory and also
for zero-mode superstring partition function, corresponding to a
rotating-moving D$p$-brane which is coupled to a $U(1)$ gauge
potential in the world-volume of the brane (photonic field) and
tachyonic background field as open string states.

By investigating the zero-modes in each case, interesting results
will be obtained. For example, according to the initial condition
and the theory (bosonic or fermionic) which is chosen, the
background fields contributions to the zero-modes would be
different. In the zero-mode bosonic boundary state, photonic
background field would be disappeared. The reason behind that can
be explained either by the boundary state method or by the path
integral techniques. The former is due to the zero-mode boundary
state equation, which is obtained by substituting the zero-mode
solution of the closed string equation of motion into the boundary
equations. It should be noted that in the case of compact
space-time this result could be changed. However, in this paper
we assume a non-compact flat space-time and therefore in the
bosonic theory, the photonic term would be absent in the zero-mode
boundary state. On the other hand, for the fermionic theory, the
electromagnetic field would be present and existence of the
tachyon field could depend on the choice of the $\sigma$'s value.

Moreover, we study the explicit form of the fermionic zero-mode
boundary state, for both type IIA and type IIB theories,
corresponding to the present rotating-moving Dp-brane with
tachyonic and photonic boundary action. By evaluating different
configurations of the described system, i.e., omitting the
background fields and/or rotating-moving structure, our system
would overlap some familiar brane structures in the literature.
In continue, we have listed some branes configurations and their
corresponding zero-mode R-R sector solutions in a classified
table. This table shows how the brane properties would affect the
R-R zero-mode solution.

Exploring the zero-mode partition function of the superstring
theory, would show the importance of the bosonic zero-mode part of
the partition function, and hence, could explain the reason of the
absence of photonic field, from the zero-mode superstring
partition function, for our described system. Besides, by
considering the low limit of the present brane structure, i.e.
omitting the dynamical movement, the familiar relation for the
zero-mode part of the effective action would be revealed.

The other achievement of this article, is obtaining an elegant
interpretation for the potential of this complex system. We show
that, by considering the rotating-moving structure and the
photonic and tachyonic fields, not only the background field but
also the dynamical structure of the system will merge to the
potential.

Simultaneous consideration of rotating-moving dynamics, and the
electromagnetic and tachyonic background fields, are the main
distinctions from conventional literature. This difference, lead
our zero-mode equations to a mixed and complex structure, between
the pointed parameters, in three cases of bosonic and fermionic
zero-mode boundary states and superstring zero-mode partition
function, and hence, make the present zero-mode investigation
outstanding.

This article is organized as follows. In Sec. 2, The zero-modes
corresponding to a rotating-moving D$p$-brane with various
background fields in the bosonic theory will be constructed. In
Sec. 3, The zero-modes in the fermionic theory of such a dynamical
system will be extracted. In Sec. 4, The zero-mode partition
function in superstring theory associated with a rotating-moving
brane will be investigated.

\section{Zero-modes in bosonic theory}

{\bf The action} - For exploring the zero-modes for a
rotating-moving D$p$-brane, the starting point is to describe
such a system by means of the sigma-model action in the bosonic
theory. The general form of the bulk action is $S =
-\frac{1}{4\pi\alpha'} {\int}_\Sigma
d^{2}\sigma(\sqrt{-h}h^{\alpha\beta}g_{\mu\nu}\partial_\alpha
X^{\mu}\partial_\beta X^{\nu})$. For the sake of simplicity,
throughout this paper we will use the conformal gauge property
and replace the two-dimensional intrinsic metric $h_{\alpha\beta}$
with the two-dimensional flat space-time metric
$\eta_{\alpha\beta}$ and use the signature
$\eta_{\mu\nu}=diag(-1,1,...,1)$ for the metric $g_{\mu\nu}$.
Therefore, the bulk action for the closed string in the conformal
gauge would be \bea S_{\rm bulk} = -\frac{1}{4\pi\alpha'}
{\int}_\Sigma d^{2}\sigma(\eta_{\mu\nu}\partial_\alpha
X^{\mu}\partial_\beta X^{\nu}), \eea  where $\Sigma$ is the
closed string world-sheet. The world-sheet is parameterized by
two parameters $(\tau,\sigma)$ and can be viewed as the map
$(\tau,\sigma)\mapsto X^{\mu}(\tau,\sigma)$. In other word, the
corresponding point on the world-sheet with coordinate
$(\tau,\sigma)$ would be $\{X^{\mu}|\mu=0,1,...,d-1\}$ on the
space-time. Besides, we consider $d^{2}\sigma\equiv d\sigma
d\tau$ and $\partial_\alpha\equiv\frac{\partial}{\partial
\sigma^{\alpha}}$ in which instead of $\sigma$ and $\tau$ one can
use variables $\sigma^{\alpha}|\alpha= 0,1$ with
$\sigma^{0}=\tau$ and $\sigma^{1}=\sigma$. Note that the
world-sheet has the topology of a cylinder which is due to the
fact that the fields $X^{\mu}$ satisfy periodic boundary
condition for the closed string, i.e.
$X^{\mu}(\tau,\sigma)=X^{\mu}(\tau,\sigma+2\pi)$ with
$0\leq\sigma\leq 2\pi$.

In continue, deformations to the original theory are coupled via
the boundary terms, as the following boundary action \bea S_{\rm
bdry}=\frac{1}{2\pi\alpha'} {\int}_{\partial\Sigma} d\sigma (
A_\alpha
\partial_{\sigma}X^{\alpha}+ \omega_{\alpha\beta}J^{\alpha\beta}_{\tau}
+iT^{2}(X^{\alpha})), \eea with the world-sheet boundary
$\partial\Sigma$. By assuming the boundary action (2), the
following background fields and related terms have been
considered: a $U(1)$ gauge field
$A_{\alpha}=-\frac{1}{2}F_{\alpha \beta }X^{\beta}$ with constant
field strength and $\partial_{\sigma}$ which is derivative along
the boundary. The second term characterizes the rotation and
motion of the brane with anti-symmetric angular velocity ${\omega
}_{\alpha \beta}$ and angular momentum density $J^{\alpha \beta
}_{\tau}$. This term is denoted by ${\omega }_{\alpha
\beta}J^{\alpha \beta }_{\tau}=2{\omega }_{\alpha \beta
}X^{\alpha }{\partial }_{\tau }X^{\beta }$, where the component
$\omega_{0 {\bar \alpha}}|_{{\bar \alpha} \neq 0}$ represents the
velocity of the brane along the direction $X^{\bar \alpha}$ and
$\omega_{{\bar \alpha}{\bar \beta}}$ which interprets its
rotation. The final term is a tachyon field with tachyonic profile
$T^{2}=T_0 +\frac{1}{2}U_{\alpha\beta}X^{\alpha}X^{\beta}$, where
$T_0$ and the symmetric matrix $U_{\alpha\beta}$ are considered
to be constant. Moreover, we represent $\{X^\alpha|\alpha =0, 1,
\cdot \cdot \cdot ,p \}$ for the world-volume directions of the
brane and $\{X^i| i= p+1, \cdot \cdot \cdot ,d-1\}$ for directions
perpendicular to it and let $U_{\alpha i }=0$ for simplicity.

{\bf The local symmetries} - As a general case, in the sigma model
approach some of the gauge symmetries of the string action on a
world-sheet with boundaries are modified by turning on the
background fields. In the above described actions (Eq. (1) and
Eq. (2)), the bulk, the background gauge field and the dynamical
term actions are diffeomorphism invariant. Besides, the tachyonic
boundary action is invariant with respect to a subgroup of the
diffeomorphism group. Therefore, Eqs. (1) and (2) are written in
world-sheet diffeomorphism invariant way. What about the Weyl
symmetry? Except the tachyonic boundary action with broken
Weyl-invariance, the other terms, i.e., the bulk, the background
gauge field and the rotating-moving term actions are Weyl
invariant and have been written in the conformal gauge. However,
despite the fact that the tachyonic boundary term does not
preserve the Weyl invariance, but still have enough symmetries to
impose the conformal gauge \cite {10}. Broken Weyl-invariance is
often referred to as broken conformal invariance and leads to an
off-shell theory. This can be expressed as an example of the
background independent string field theory. In this content, one
can study properties of the unstable D-branes due to the tachyon
condensation (where the unstable D-branes may behave like
solitons, corresponding to lower dimensional D-branes) \cite
{8}-\cite {10}. The effect of losing the conformal invariance,
due to the presence of tachyon field, would show up not only as
corrections to the effective action \cite {12} but also in the
boundary states. At the end of this section, we will come back to
this point and discuss the effect of this conformal breaking on
the boundary states of the described system.

{\bf Boundary state} - Boundary state equations for the above
bosonic action can be achieved by vanishing the variation of the
action with respect to $X^{\mu }(\sigma ,\tau )$ as
 \bea &~& [{(\eta }_{\alpha \beta }
+4{\omega }_{\alpha \beta }){\partial }_{\tau }X^{\beta }
+{F_{{\mathbf \alpha \beta}}}{\partial }_{\sigma }X^{\beta }
+iU_{\alpha \beta }X^{\beta }]|_{\tau =0}\ \ {|B\rangle}_{bos} =0,
\nonumber\\
&~& {\delta X}^i|_{\tau =0}{|B\rangle}_{bos} =0, \eea This
equation is the key relation for obtaining the zero mode boundary
states in both bosonic and fermionic sections.

{\bf Bosonic zero-mode boundary state} - As a general strategy
for the bosonic theory, the zero-mode coordinate $X$ is conjugate
to the total momentum $p$ of the string. In the absence of an
external field, the zero-mode boundary-state is the
translation-invariant state ${|B\rangle}_{bos}^{(0)}$ annihilated
by $p$. In the case of having a boundary action $S(X .... )$
which depends on $X$ through a space-dependent background field,
this zero-mode state is changed by the action of the zero-mode
coordinate, to $e^{-S(q)}{|B\rangle}_{bos}^{(0)}$ \cite {4}.

Now, let us find the zero-mode bosonic boundary state, first, by
applying the zero-mode closed string mode expansion \bea &~&
X^{\mu }(\sigma ,\tau )=x^{\mu }+l^2p^{\mu }\tau
+\frac{1}{2}il\sum_{m\ne 0}{\frac{1}{m}\ (\ {\alpha }^{\mu }_m
e^{-2im\left(\tau -\sigma \right)} +}{\widetilde{\alpha }}^{\mu
}_me^{-2im(\tau +\sigma )}), \nonumber \eea in Eqs. (3).
Therefore, in the case of our background fields, boundary state
equations  are converted to \bea [2\alpha'({\eta }_{\alpha \beta
}+4{\omega }_{\alpha \beta })p^{\beta } +iU_{\alpha \beta
}x^{\beta }]{|B\rangle}_{bos}^{(0)} =0, \eea \bea
(x^{i}-y^{i}){|B\rangle}_{bos}^{(0)} =0, \eea Where the set
$\{y^i\}$ indicates the position of the brane.

Then, by combining Eqs. (4) and (5) and using the following
relation \bea |B\rangle\ =\int^{\infty }_{{\rm -}\infty }
{\prod_{\alpha }{dp^{\alpha }} \langle{p^{\alpha} |B\rangle\ \
|p^{\alpha }\rangle\ }} \eea the zero-mode bosonic boundary state
is obtained as \bea {|B\rangle}_{bos}^{(0)} &=& \int^{\infty
}_{{\rm -}\infty } {\prod_{\alpha }{dp^{\alpha
}}}\exp\left[{\alpha }^{{\rm '}}\left(
\sum^p_{\alpha=0}{\left(U^{{\rm -}{\rm 1}}{\mathbf
A}\right)}_{\alpha \alpha } {\left(p^{\alpha }\right)}^{{\rm
2}}{\rm +} \sum^{p}_{\alpha ,\beta {\rm =0},\alpha \ne
\beta}{{\left(U^{{\rm -}{\rm 1}}{\mathbf A}+{\mathbf A}^T
U^{-1}\right)}_{\alpha \beta } p^{\alpha }p^{\beta
}}\right)\right]{\rm \ \ }
\nonumber\\
&\times& \prod_i{\delta {\rm (}x^i}{\rm -}y^i{\rm )} {\rm
|}p^i{\rm =0}\rangle \otimes \prod_\alpha {\rm |}p^{\alpha
}\rangle . \eea where ${\mathbf A}_{\alpha \beta}=\eta_{\alpha
\beta} + 4\omega_{\alpha \beta}$.

The above momentum dependent integral comes from taking into
account the first equation of zero modes, i.e, Eq. (4) and a
delta function, which is inserted to fix the location of the
D-brane (by imposing an extra condition on the position operator)
in the transverse direction according to Eq. (5).

It should be noted that for simplifying the calculations, the
Kalb-Ramond field $B_{{\mu}{\nu}}$ has been omitted from the bulk
action. However, even if one considered this field in addition to
the gauge potential, then, the results of the zero-mode would not
be changed. Because it could be converted into the boundary term
by defining ${\cal{F}}_{\alpha \beta}=F_{\alpha \beta}-B_{\alpha
\beta}$. Therefore, addition of this term does not have any
contribution to the zero-mode bosonic boundary state.
Consideration of the bulk 2-form $B$ and the field strength $F$,
would show up only in the oscillating part of the bosonic
boundary state.

There is an interesting point that except the photonic field, the
other background fields have been contributed in the zero mode
boundary state of the bosonic theory. The reason behind that
refers to Eq. (4) and the fact that zero mode closed string mode
expansion does not contribute to the photonic field. It's
physical point of view can be explained as follows: in the case
of static brane, the tachyon does not couple to the gauge field
because the tachyon is in the U(1) adjoint representation, hence,
using the path integral method and considering this decoupling
from the gauge field, the zero-mode would just depends on tachyon
zero mode \cite {9}. in the case of adding a dynmical structure
to the same brane, the mentioned decoupling of the gauge field
would be exist and the rotating-moving feature of the brane would
be appear along with the tachyon field in the zero-mode bosonic
boundary state. Thus, inclusion of rotating-moving dynamics of
the Dp-brane, which is emerged from the momentum dependent
integral, would be create an outstanding mixed structure for the
tachyon and this special dynamics of the system. It should be
noted that in the case of compact space-time, one could reach to
a different result. This is due to the fact that for compact
directions, the complete solution of the closed string equation
has a $2L^{\mu}\sigma$ term, proportional to the radius of
compactification in the compact direction. For non-compact
directions, the closed string winding number $L^{\mu}$ is zero but
in the case of having components along the compact directions
this term would be remained. Therefore, because of the $\sigma$
dependency of the photonic term in the bosonic boundary state
equation, all of the background fields could be contributed in
the zero mode boundary state of the bosonic theory.

Now let us point out to an important note here. As we know, due to
the presence of the tachyon field in the boundary state, conformal
invariance has been broken. However, we know that D-branes are
specified by conformally invariant boundary states which act as
sources for all closed string fields. Therefore, in order to have
a conformal invariant boundary state, one could deform it under
the subset of the full conformal group of the plane, PSL(2,R),
which preserves the position and shape of the boundary of the
disc. In other words, the loss of conformal invariance introduced
by the background tachyon feld is accomodated by a conformal
transformation which induces a calculable change in the boundary
state. Obtaining the new modified boundary state one could see
the fact that the zero mode part of the boundary state does not
affected by these transformations. The detailed calculations can
be found in \cite {7}. For the sake of simplicity, we do not use
the modified form of the boundary state here.

These interesting features caused the present zero-mode bosonic
boundary state to be distinguished from the conventional
literature.

\section{Zero-modes in fermionic theory}

The supersymmetric prescription of the bosonic action, in the
previous section, is invariant under the global world-sheet
supersymmetry. Therefore, one can find the fermionic partners of
the bosonic boundary equations by applying the supersymmetry
transformations. According to the bosonic boundary state
equations (3) we need to insert the following replacements \bea
{\partial }_{\sigma }X^{\mu }\left(\sigma ,\tau \right)\to
(-{\psi }^{\mu }_{0}-i\eta {\tilde{\psi }}^{\mu }_{0})+\sum_{k\ne
0} {(-}i\eta {\tilde{\psi }}^{\mu }_{k\ }\ e^{-2ik(\tau +\sigma
)}-{\psi }^{\mu }_{k\ }\ e^{-2ik(\tau -\sigma )})
\nonumber\\
{\partial }_{\tau }X^{\mu }\left(\sigma ,\tau \right)\to ({\psi
}^{\mu }_{0}-i\eta {\tilde{\psi }}^{\mu }_{0})+\sum_{k\ne 0}
{(-}i\eta {\tilde{\psi }}^{\mu }_{k\ }\ e^{-2ik(\tau +\sigma
)}+{\psi }^{\mu }_{k\ }\ e^{-2ik(\tau -\sigma )}) \eea These two
general forms for supersymmetry transformations have been
calculated by considering the world-sheet supersymmetry
transformations \bea {\partial }_+X^{\mu }\left(\sigma ,\tau
\right)\to -i\eta {\psi }^{\mu }_ +\left(\sigma ,\tau \right),
\nonumber\\
{\partial }_-X^{\mu}\left(\sigma ,\tau \right)\to {\psi }^{\mu
}_-\left(\sigma ,\tau \right),\eea ${\partial }_{\pm }=({\partial
}_{\tau }\pm {\partial }_{\sigma })/2$ and the solution of the
equations of motion for the fermions. In addition $\eta = ±1$ is
due to the GSO projection of the boundary state.

Furthermore, to fulfill applying the supersymmetry transformations
of Eq. (3), we need the following replacement for the tachyonic
term \bea X^{\beta }\left(\sigma ,\tau \right)\to \sum_{k\ne
0}{\frac{1}{2ik}(}i\eta {\tilde{\psi }}^{\beta }_{k\ }\
e^{-2ik(\tau +\sigma )} -{\psi }^{\beta }_{k\ }\ e^{-2ik(\tau
-\sigma )})+(-i\eta {\tilde{\psi }}^{\beta }_{0}-{\psi }^{\beta
}_{0})\sigma+({\psi }^{\mu }_{0}-i\eta {\tilde{\psi }}^{\mu
}_{0})\eea

Note that it is important to separate the zero modes and non-zero
modes in Eqs. (8) and (10) to prevent having a wrong answer. The
significance of this point will be revealed in the zero-mode
fermionic boundary equations.

Considering all these together, the zero-mode fermionic boundary
states for the closed string boundary at $\tau=0$ are as \bea &~&
[({\eta }_{\alpha \beta }+4{\omega }_{\alpha \beta })(-i\eta
{\tilde{\psi }}^{\beta }_{0}+{\psi }^{\beta }_{0})+ (F_{\alpha
\beta }+i\sigma U_{\alpha \beta })(-i\eta {\tilde{\psi }}^{\beta
}_{0}-{\psi }^{\beta }_{0})]{|B\rangle}_{ferm}^{(0)} =0
\nonumber\\
&~&(-i\eta {\tilde{\psi }}^{i}_{0}-{\psi
}^{i}_{0}){|B\rangle}_{ferm}^{(0)} =0. \eea The two Eqs. (11) can
be collected into a unified form \bea (d^{\mu}_{0}-i\eta
S^{\mu}_{(0)\nu}\tilde{d}^{\nu}_{0}){|B\rangle}_{R-R}^{(0)} =0
\eea in which \bea &~& S_{(0)\mu\nu}=(\Delta_{(0)\alpha
\beta},-\delta_{ij}),
\nonumber\\
&~& \Delta_{(0)\alpha \beta}=(M_{(0)}^{-1}N_{(0)})_{\alpha \beta},
\nonumber\\
&~& M_{(0)\alpha \beta}=({\eta }_{\alpha \beta }+4{\omega
}_{\alpha \beta }-F_{\alpha \beta }-i\sigma U_{\alpha \beta }),
\nonumber\\
&~& N_{(0)\alpha \beta}=({\eta }_{\alpha \beta }+4{\omega
}_{\alpha \beta }+F_{\alpha \beta }+i\sigma U_{\alpha \beta }).
\eea

Clearly the only contribution of the fermionic zero-mode part of
the boundary state is that of the R-R sector. It is due to the
mode number in the NS-NS sector which runs over half integers.

An important point here is, unlike the bosonic zero-mode boundary
state in which photon was omitted, in the fermionic case all the
background fields could be participated. This matter depends on
the value of the $\sigma$. Therefore in the case of $\sigma=0$
the contribution of tachyonic field would be disappeared and the
tachyon would be omitted from the zero-mode boundary state. The
final result would affects the fermionic vacuum and hence the
spin structure of the boundary state.

To complete our discussion, we should find the explicit form of
the fermionic zero-mode boundary state $|B\rangle_{ferm}^{(0)}$
by applying similar considerations as in the bosonic section.
According to \cite {4}, the general strategy is as follows: In
the presence of a non vanishing background field, the zero-mode
part of the boundary action, will generate some linear
combination of n-forms, that span a $2^d$-dimensional Hilbert
space. This work is done by the action of the zero-mode part on
the vacuum. Therefore, the zero-mode part can be classified as
\bea &~& e^{-S(\theta^{\mu}_{0}, .... )}|0;+\rangle = \{\rm
polynomial \ in \ \theta^{\mu}_{0}\} |0;+\rangle,\nonumber \eea
for the $+$ spin structure, and \bea &~& e^ {-S(\pi^{\mu}_{0},
.... )}|0;-\rangle = \{\rm polynomial \ in \ \pi^{\mu}_{0}\}
|0;-\rangle,\nonumber \eea for the $-$ spin structure. In these
relations $\theta^{\mu}_{0}$ are the boundary coordinates
associated with the zero modes of the $+$ spin structure with the
definition
$\theta^{\mu}_{0}=\psi^{\mu}_{0}+i{\tilde{\psi}}^{\mu}_{0}$, and
$\pi^{\mu}_{0}$ are their corresponding conjugate momenta defined
as $\pi^{\mu}_{0}=1/2(\psi^{\mu}_{0}-i{\tilde{\psi}}^{\mu}_{0})$.

The vacuum of one spin structure is the filled Fermi sea of the
other, due to the fact that creation operators of one spin
structure are the annihilation operators of the other, i.e.,
$|0;-\rangle=\prod_{\mu=1}^{D}{\theta^{\mu}_{0}|0;+\rangle}$. The
states in the zero-mode Hilbert space can be represented by
antisymmetrized products of D-dimensional gamma matrices, this is
due to the fact that the n-forms (built out of the
$\theta^{\mu}_{0}$), correspond to the D-dimensional Clifford
algebra. Now by considering a duality relation between the above
zero-mode forms, and transforming the antisymmetrized products of
gamma matrices (by multiplying them by $\Gamma_{11}$), the gamma
matrix representation of the zero-mode forms can be adopted as
\bea &~& e^{-S(\theta^{\mu}_{0}, .... )}|0;+\rangle \equiv \{\rm
sum \ of \ \Gamma^{{\mu}_{1}...{\mu}_{n}}\} |0;+\rangle,
\nonumber\\
&~& e^{-S(\pi^{\mu}_{0}, .... )}|0;-\rangle \equiv \{\rm sum \ of
\ \Gamma^{{\mu}_{1}...{\mu}_{n}}\}\Gamma_{11} |0;+\rangle \eea

Therefore, by unifying these two spin structures provided by the
gamma matrix notation, the products of fermionic zero modes can be
represented as antisymmetrized products of gamma matrices, along
with introducing a notation in which all terms in the expansion
with repeated Lorentz indices are to be dropped.

Using above considerations, let us study the explicit form of the
fermionic zero-mode boundary state $|B\rangle_{ferm}^{(0)}$ both
in type IIA and type IIB theories, for the present case of a
rotating-moving Dp-brane with tachyonic and photonic boundary
action. Considering the spin fields in the 32-dimensional Majorana
representation, i.e, ${\textbf{S}}^{A}$ and
$\tilde{{\textbf{S}}}^{B}$, the vacuum for the fermionic
zero-modes has the following feature \cite {5} \bea
|A\rangle|\tilde{B}\rangle=\lim_{z,\bar{z}\rightarrow 0}
{\textbf{S}}^{A}(z){\tilde{{\textbf{S}}}}^{B}(\bar{z})|0\rangle
\eea Where $A,B=1,...,32$.

Translating above considerations into Hilbert space language, the
action of $\psi^{\mu}_{0}$ and $\tilde{\psi}^{\mu}_{0}$ on the
R-R vacuum, in a non-chiral basis, can be obtained as \bea
&~&\psi^{\mu}_{0}|A\rangle|\tilde{B}\rangle=\frac{1}{\sqrt{2}}{(\Gamma^{\mu})^{A}_{C}}{(1)^{B}_{D}}|C\rangle|\tilde{D}\rangle
\nonumber\\
&~&
\tilde{\psi}^{\mu}_{0}|A\rangle|\tilde{B}\rangle=\frac{1}{\sqrt{2}}{(\Gamma_{11})^{A}_{C}}{(\Gamma^{\mu})^{B}_{D}}|C\rangle|\tilde{D}\rangle
\eea

Using these definitions, the zero-mode fermionic structure of the
present system, i.e, $|B\rangle_{ferm}^{(0)}$, in the language of
Hilbert space, can be obtained by considering a solution of Eq.
(12) as \bea
|B\rangle_{ferm}^{(0)}=\Lambda_{AB}|A\rangle|\tilde{B}\rangle \eea
in which $\Lambda$ satisfies $(\Gamma^{\mu})^{T}\Lambda-i\eta
S^{\mu}\Gamma_{11}\Lambda \Gamma^{\nu}=0 $, where a chiral
representation for the $32\times32$ $\Gamma$-matrices is applied.

Consequently, the following solution can be found  \bea \Lambda=C
\Gamma^{0}\Gamma^{1}...\Gamma^{p}(\frac{1+i\eta\Gamma_{11}}{1+i\eta})\Upsilon
\eea where $C$ is the charge conjugate matrix, and \bea
\Upsilon=K\diamondsuit \exp\left\{\frac{1}{2}[(
\Delta_{(0)}-1)(\Delta_{(0)}+1)^{-1}]_{\alpha\beta}\Gamma^{\alpha}\Gamma^{\beta}\right\}\diamondsuit,
\eea in which $K$ is a normalization constant that should be
determined according to the boundary action and total boundary
state, including the oscillating part of the both bosonic and
fermionic sections. Here we do not intend to calculate it. The
interested reader can find the detailed calculations in our
previous papers \cite {6}. Moreover, the symbol $\diamondsuit$
rules that the exponential should be expanded and the indices of
the $\Gamma$-matrices must be antisymmetrized. As a result, this
symbol makes some restrictions. For example, it limits the number
of terms for each value of $p$ to a finite number.

Now let's see the above fermionic zero-mode solution how could be
changed with some simplifications. Note that the general solution
would be the same form as Eq. (17), but due to the initial
assumptions, $\Lambda_{AB}$ would be different. First, consider
the Dp-brane as the static object without any background fields.
For this structure the fermionic zero-mode solution would be as $
\Lambda=C
\Gamma^{0}\Gamma^{1}...\Gamma^{p}(\frac{1+i\eta\Gamma_{11}}{1+i\eta})$,
in which we have $ \Upsilon=1 $. Now by considering the
excitations of the open strings attached to the Dp-brane (adding
gauge field) $ \Lambda $ would be change according to
$\Upsilon=\sqrt{\det(\eta-F)}\diamondsuit
\exp\left\{\frac{1}{2}F_{\alpha\beta}\Gamma^{\alpha}\Gamma^{\beta}\right\}\diamondsuit.$
These two low limits of the present structure are in complete
agreement with the literature. It should be noted that according
to the initial definition of our gauge filed the prefactor $K$ is
the inverse value of \cite {5}.

In table 1  some branes configurations and their corresponding
zero-mode R-R sector solutions have been listed. It should be
noted that the T-duality transformation mentioned in the list is
performed in the transverse direction $i^{th}$ of the D-brane.
And for the next case, at first a rotation in the $(p, p + 1)$
plane with angle $\phi$ and then a T-duality transformation in
the $(p + 1)^{th}$ direction have been done. The rotated and
T-dualized boundary state is the world-sheet realization of a
bound state which reproduces the long-distance behavior of the
delocalized D-brane classical solutions \cite {5}.

As it seems, the present system can cover some of those
configurations in the low limits (omitting the background fields
and/or rotating-moving structure). This outstanding feature is
appeared also in the zero-mode partition function in superstring
theory.

\pagebreak

\begin{tabular}{ | p{2cm} | c | p{12cm} |}
    \hline
    \textit{\textbf{Brane configuration}} & \textit{\textbf{R-R zero-mode solution}}  \\ \hline

    \footnotesize Static Brane, No external field & $ \Lambda=C
\Gamma^{0}\Gamma^{1}...\Gamma^{p}(\frac{1+i\eta\Gamma_{11}}{1+i\eta})
$ \\ \hline
     \footnotesize T-dualized Brane, No external field & $\Lambda=C
\Gamma^{0}\Gamma^{1}...\Gamma^{p}\Gamma^{i}(\frac{1+i\eta\Gamma_{11}}{1+i\eta})$
 \\ \hline
     \footnotesize Rotated and T-dualized Brane, No external field & $\Lambda=C
\Gamma^{0}\Gamma^{1}...\Gamma^{p-1}(\sin \phi+\cos\phi
\Gamma^{p}\Gamma^{p+1}) $
 \\ \hline

    \footnotesize Boosted Brane, No external field & $\Lambda=\frac{1}{\sqrt{1-v^2}}(C[
\Gamma^{0}+v\Gamma^{k}]\Gamma^{1}...\Gamma^{p})(\frac{1+i\eta\Gamma_{11}}{1+i\eta})$
\\ \hline
     \footnotesize Static Brane, With gauge field & $\Lambda=\sqrt{\det(\eta-F)}C
\Gamma^{0}\Gamma^{1}...\Gamma^{p}(\frac{1+i\eta\Gamma_{11}}{1+i\eta})\diamondsuit
e^{\frac{1}{2}F_{\alpha\beta}\Gamma^{\alpha}\Gamma^{\beta}}\diamondsuit$
\\ \hline
    \footnotesize Rotating-Moving Brane, With F and U fields & $\Lambda=\sqrt{\det(\eta -F)}\det \left[\frac{\sqrt{\pi}\Gamma \left(1-\frac{1}{2}(\eta -F)^{-1}U \right)}
    {\Gamma\left(\frac{1}{2}-\frac{1}{2}(\eta -F)^{-1}U \right)
    }\right] C\Gamma^{0}\Gamma^{1}...\Gamma^{p}(\frac{1+i\eta\Gamma_{11}}{1+i\eta})\diamondsuit
     e^{\frac{1}{2}[\frac{\Delta_{(0)}-1}{\Delta_{(0)}+1}]_{\alpha\beta}\Gamma^{\alpha}\Gamma^{\beta}}\diamondsuit$
 \\ \hline

\end{tabular}

\begin{table}[htbp]\centering
\def\sym#1{\ifmmode^{#1}\else\(^{#1}\)\fi}
 \caption{\footnotesize some branes configurations and
their corresponding zero-mode R-R sector solutions.}
\end{table}

\section{Interpretation of zero-mode partition function in superstring theory}

According to the description of the boundary string field theory,
a fixed conformal world-sheet action in the bulk (in the space of
two dimensional world-sheet theories on the disk) with arbitrary
boundary deformations, construct a configuration space for the
boundary string field theory. In this context, propagating a
closed string from the boundary of the disk can be explained as
the disk partition function in the closed string theory. It has
been demonstrated that the vacuum amplitude of the boundary state
can be described as the partition function \bea
Z^{Disk}_{S_{bdry}}=\langle vacuum|B;S_{bdry}\rangle. \eea

Now by inserting both the bosonic and fermionic zero-mode boundary
states obtained in the previous sections, integrating on the
momenta and projecting that onto the bra-vacuum, the zero-mode
part of the partition function in superstring theory would be
appeared as \bea Z^{(0)}_{Super}=\left(-\frac{\pi}{\alpha'}
\right)^{(p+1)/2} \frac{1}{\sqrt{{\det (H + D)}}},\eea where
$D_{\alpha \beta}= (U^{-1}{\mathbf A})_{\alpha
\alpha}\delta_{\alpha \beta}$ and \bea H_{\alpha \beta}=
\bigg{\{} \begin{array}{c}
(U^{-1}{\mathbf A} + {\mathbf A}^T U^{-1})_{\alpha \beta}\;\;,\;\;\;\alpha \neq \beta ,\\
0 \;\;\;\;\;\;\;\;\;\;\;\;\;\;\;\;\;\;\;\;\;\;\;\;\;\;\;\;\;\;
\;,\;\;\;\alpha = \beta,
\end{array}.\eea

Note that the NS-NS and R-R sectors of the fermionic boundary
states have contribution in the partition function but not in the
zero-mode part, hence, they would not appear here. Moreover, the
explicit form of the fermionic zero-mode state
$|B\rangle_{ferm}^{(0)}$ and its contribution to the spin
structure have been projected out, according to the above
description, and therefore omitted from the zero-mode partition
function in superstring theory. Thus, the significant part of the
zero-mode partition function is that of the bosonic one.

As it seems, the photonic field has been disappeared. The reason
behind this can be explained in two different ways: First, this
is due to the fact that the tachyon is in the $U(1)$ adjoint
representation, hence, it could not couple to the gauge field.
Second, it can be explained by applying the boundary state method,
in which by making use of the bosonic zero-mode boundary state
(7), the photonic term would not be present any more.

A complex mixture of tachyonic background field and the
rotating-moving dynamics of the system, distinguish this equation.
In order to see this issue in detail, let us calculate the
explicit form of the symmetric matrix $\Phi=H+D$ which is the
heart of our zero-mode part of the superstring partition
function. For this purpose, consider a $D2$-brane with the linear
velocity $v_{\bar \alpha}$ where $\alpha$ could get the values of
1, 2, and the angular velocity $(\omega_{12}) = \overline{\omega
}$. Then the matrix elements of $\Phi$ would be \bea &~&
\Phi_{00}=(U^{-1})_{00}-4v_1(U^{-1})_{01}-4v_2(U^{-1})_{02}\;,
\nonumber\\
&~& \Phi_{01}=
2(U^{-1})_{01}-4v_1\left((U^{-1})_{00}+(U^{-1})_{11}\right)
-4\overline{\omega }(U^{-1})_{02}-4v_2(U^{-1})_{21}\;,
\nonumber\\
&~&
\Phi_{02}=2(U^{-1})_{02}-4v_2\left((U^{-1})_{00}+(U^{-1})_{22}\right)
+4\overline{\omega }(U^{-1})_{01}-4v_1(U^{-1})_{12}\;,
\nonumber\\
&~& \Phi_{11}=(U^{-1})_{11}-4v_1(U^{-1})_{10} -4\overline{\omega
}(U^{-1})_{12}\;,
\nonumber\\
&~& \Phi_{12}= 2(U^{-1})_{12}+4\overline{\omega }
\left((U^{-1})_{11}-(U^{-1})_{22}\right)
-4v_2(U^{-1})_{10}-4v_1(U^{-1})_{02}\;,
\nonumber\\
&~& \Phi_{22}=(U^{-1})_{22}-4v_2(U^{-1})_{20} +4\overline{\omega
}(U^{-1})_{21}. \eea Clearly, a complicated structure between the
elements of the tachyon background field, rotation and motion of
the brane can be seen from the above relations which distinguish
our work from the others where $U_i$s do not mix with any other
couplings.

In continue let's introduce a new interpretation for the
zero-modes of the present superstring partition function, by
concentrating on the effective action. In order to reach this
concept first, consider the familiar tachyon effective action with
photonic term in superstring theory \cite {8,13} \bea
\textit{L}=-V(T)\sqrt{-\det(\textit{G}_{\mu\nu}+\textit{F}_{\mu\nu})}
\textit{K}((\textit{G}+\textit{F})^{\mu\nu}\partial_{\mu}T\partial_{\nu}T).\eea
Then look at the effective action associated with our
rotating-moving D$p$-brane with background fields extracted from
the boundary string field theory \cite {12} \bea \textit{L}'=
-V(T)\frac{T_p}{g_s} \frac{\left(-\frac{\pi}{\alpha'}
\right)^{(p+1)/2}}{\sqrt{{\det (H + D)}}}\sqrt{
-\det(\eta+F)}\textit{K}'((\eta+\textit{F})^{\mu\nu}\partial_{\mu}T\partial_{\nu}T)\eea
Where the tachyonic potential term $V(T)$ in superstring field
theory action is proportional to $e^{-T^2/4}$ due to the tachyon
profile and $\textit{K}$ and $\textit{K}'$ are two general
functionals according to the considered system.

By comparing the effective action (24), (static brane with
tachyonic and photonic background fields) and the generalized
form of our Lagrangian in (25), (rotating-moving brane with the
same fields), the potential term of the present dynamical
D$p$-brane with photonic and tachyonic background fields can be
obtained as \bea V_{\textit{system}}=
\frac{T_p}{g_s}V(T)Z_{Super}^{(0)},\eea An interesting point here
is that the dynamics of this D$p$-brane (rotating-moving
movement) is just inserted into the system's potential term and
not in any other part of the effective action. This matter can be
explained either by the boundary state equations or the path
integral calculations. Therefore, unlike the the former case in
which the only argument of the potential is the zero-mode of the
tachyon field, the potential of this dynamical D$p$-brane is
proportional to the zero-mode part of the superstring partition
function, including a mixture of tachyon and rotating-moving
dynamics. Hence, in addition to the tachyonic field, the
dynamical structure of this system also is merged to the
potential.

Now let us investigate the zero-mode partition function of the
described system for the static case, i.e, in the absence of
$\omega$-term. This can be done by setting $\omega=0$ in the
above relations. But it should be noted that we calculated these
equations in the momentum space, then, let us return back and
calculate the zero-mode partition function in the $x$-space as
\bea Z^{(0)}=(-4\pi\alpha')^{\frac{p+1}{2}} \frac{1}{\sqrt{{\det
U}}},\eea This is the familiar relation for the zero-mode part of
the effective action in \cite {9, 14}.


\end{document}